\documentclass[twocolumn,secnumarabic,amssymb,superscriptaddress,nobibnotes, aps, pra]{revtex4-1}

\usepackage{amscd}
\usepackage{color}
\usepackage{graphicx}

\newcommand{\beq}{\begin{eqnarray}}
\newcommand{\eeq}{\end{eqnarray}}
\newcommand{\im}{{i}}

\newcommand{\del}{\partial}
\newcommand{\de}{\mathrm{d}}

\newcommand{\e}{\mathrm{e}}

\newcommand{\A}{\mathbf{A}}
\newcommand{\B}{\mathbf{B}}

\newcommand{\E}{\mathbf{E}}

\newcommand{\pp}{\mathbf{p}}
\newcommand{\xx}{\mathbf{x}}
\newcommand{\rr}{\mathbf{r}}

\newcommand{\PP}{\mathbf{P}}

\newcommand{\C}{\mathbf{C}}

\begin{document}

\title{Gauge Invariance beyond the Electric Dipole Approximation}

\begin{abstract}
We study the gauge invariance of laser-matter interaction. 
The velocity gauge where the vector potential is expanded to the $n$-th order with respect to the spatial coordinate, and the length gauge where the electric and magnetic fields are expanded to the $n$-th and $(n-1)$-th orders, respectively, are mutually gauge-transformed, describing the physically equivalent situation. 
The latter includes up to the electric $2^{n+1}$-pole and magnetic $2^n$-pole interactions as well as two extra terms.
The finding serves to develop consistent nonperturbative simulation methods beyond the electric dipole approximation.
\end{abstract}

\author{Ryoji Anzaki}
	\email[]{anzaki@atto.t.u-tokyo.ac.jp}
	\affiliation{Department of Nuclear Engineering and Management, Graduate School of Engineering, The University of Tokyo, 7-3-1 Hongo, Bunkyo-ku, Tokyo 113-8656, Japan}
\author{Yasushi Shinohara}
    \affiliation{Photon Science Center, Graduate School of Engineering, The University of Tokyo, 7-3-1 Hongo, Bunkyo-ku, Tokyo 113-8656, Japan}
\author{Takeshi Sato}
  	\affiliation{Department of Nuclear Engineering and Management, Graduate School of Engineering, The University of Tokyo, 7-3-1 Hongo, Bunkyo-ku, Tokyo 113-8656, Japan}
    \affiliation{Photon Science Center, Graduate School of Engineering, The University of Tokyo, 7-3-1 Hongo, Bunkyo-ku, Tokyo 113-8656, Japan}
  	\affiliation{Research Institute for Photon Science and Laser Technology, The University of Tokyo, 7-3-1 Hongo, Bunkyo-ku, Tokyo 113-0033, Japan}
\author{Kenichi L. Ishikawa}
  	\affiliation{Department of Nuclear Engineering and Management, Graduate School of Engineering, The University of Tokyo, 7-3-1 Hongo, Bunkyo-ku, Tokyo 113-8656, Japan}
    \affiliation{Photon Science Center, Graduate School of Engineering, The University of Tokyo, 7-3-1 Hongo, Bunkyo-ku, Tokyo 113-8656, Japan}
  	\affiliation{Research Institute for Photon Science and Laser Technology, The University of Tokyo, 7-3-1 Hongo, Bunkyo-ku, Tokyo 113-0033, Japan}

\date{\today}

\maketitle

\section{Introduction}
\label{sec:introduction}

The electron dynamics in atoms and molecules subject to ultrashort intense (visible-to-near-infrared) laser pulses and extreme-ultraviolet pulses are widely simulated on the basis of the time-dependent Schr\"odinger equation.
Besides the single-active-electron (SAE) approximation, various {\it ab initio} multielectron methods \cite{Ishikawa2015JSTQE} have been developed, such as time-dependent close-coupling \cite{Parker1996JPB}, time-dependent configuration-interaction singles \cite{Rohringer2006,Greenman2010PRA,Sato2018AS}, time-dependent $R$-matrix \cite{Lysaght2008PRL}, time-dependent multiconfiguration self-consistent-field (with time-varying orbitals) \cite{Zanghellini2003LP,Kato2004CPL,Caillat2005PRA,Sato2013PRA,Sato2015PRA,Sato2016PRA,Sawada2016PRA,Anzaki2017PCCP,Miyagi2013PRA,Miyagi2014PRA,Bauch2014PRA,Haxton2015PRA}, time-dependent coupled-cluster \cite{Kvaal2012,Sato2018JCP}, time-dependent algebraic diagrammatic construction \cite{Ruberti2018PCCP}, time-dependent density-functional-theory \cite{Yabana2012PRB}, time-dependent two-particle reduced-density-matrix \cite{Lackner2015PRA,Lackner2017PRA}, and state-specific expansion methods \cite{Mercouris1994PRA}.
Their vast majority uses the electric dipole approximation (EDA), within which it is known that the laser-electron interaction is expressed either in the length gauge (LG) or velocity gauge (VG) \cite{bandrauk2013}.
In the SAE case, for example, the LG and VG Hamiltonians read (we use Hartree atomic units unless otherwise stated),
\beq
\label{eq:intro-LG0} H_{\rm LG}^{\rm (E1)} &=& \frac{{\bf \hat{p}}^2}{2} + \rr\cdot\E(0,t) + V_{\rm eff}({\bf r}),\\
\label{eq:intro-VG0} H_{\rm VG}^{\rm (E1)} &=& \frac{[{\bf \hat{p}}+\A(0,t)]^2}{2}  + V_{\rm eff}({\bf r}),
\eeq 
respectively, where $\E(\rr,t)=-\dot{\A}(\rr,t)$ denotes the electric field, $\A(\rr,t)$ the vector potential, and $V_{\rm eff}$ the effective potential.
A unitary operator $W_0 = \e^{- \im \rr\cdot\A(0,t)}$ bridges between the LG and VG Hamiltonians:
\beq
{H}_{\rm LG}^{\rm (E1)} &=& W_0^{-1}{H}^{\rm (E1)}_{\rm VG}W_0 - \im W_0^{-1}\dot{W}_0,\\ \quad {H}^{\rm (E1)}_{\rm VG} &=& W_0{H}^{\rm (E1)}_{\rm LG}W_0^{-1} - \im W_0\dot{W}_0^{-1}.
\eeq
Similarly, the LG and VG wave functions $\psi_{\rm LG}^{\rm (E1)}({\bf r},t)$ and $\psi_{\rm VG}^{\rm (E1)}({\bf r},t)$, respectively, are related via $\psi_{\rm VG}^{\rm (E1)}({\bf r},t)=W_0\psi_{\rm LG}^{\rm (E1)}({\bf r},t)$.
The gauge principle, one of the fundamental principles in modern physics, states that all physical observables are gauge invariant, i.e., take the same values whether the length or velocity gauge may be used \cite{bandrauk2013}.

Recent development in circularly and elliptically polarized high-harmonic generation (HHG) \cite{fleischer2014} has triggered growing interest in the optical response of chiral molecules, e.g., photoelectric circular dichroism \cite{dreissigacker2014} and detection of enantiomers \cite{cireasa2015,Harada2018PRA,smirnova2015, wong2011, patterson2013, fanood2015}.
To simulate chiral molecules, we have to include the magnetic dipole interaction \cite{autschbach2012}.
For the LG case, the extension of $H_{\rm LG}^{\rm (E1)}$ is straightforward:
\begin{equation}
	\label{eq:intro-E1+M1}
	H_{\rm LG}^{\rm (E1+M1)} = \frac{{\bf \hat{p}}^2}{2} + \rr\cdot\E(0,t) + \frac{\rr\times{\bf \hat{p}}}{2}\cdot\B(0,t) + V_{\rm eff}({\bf r}),
\end{equation}
which has been used, e.g., in Ref.~\cite{cireasa2015}.
On the other hand, if one has been using VG for simulations within EDA, its natural extension may be expansion of ${\bf A}({\bf r},t)$ to the first order in ${\bf r}$:
\begin{equation}
	\label{eq:intro-VG1}
	H_{\rm VG}^{\rm (1)} = \frac{[{\bf \hat{p}}+\A(0,t)+(\rr\cdot\del_\xx)\A(\xx,t)\big|_{\xx=0}]^2}{2}   + V_{\rm eff}({\bf r}),
\end{equation}
or that of $[{\bf \hat{p}}+\A({\bf r},t)]^2$ to the first order in ${\bf r}$:
\begin{eqnarray}\nonumber
	\label{eq:intro-VG1prime}
	H_{{\rm VG}^\prime}^{\rm (1)} &=& \frac{[{\bf \hat{p}}+\A(0,t)]^2}{2}\\ 
	&+& (\rr\cdot\del_\xx)\A(\xx,t)\big|_{\xx=0}\cdot[{\bf \hat{p}}+\A(0,t)] + V_{\rm eff}({\bf r}).
\end{eqnarray}
More generally, one can further extend Eq.~(\ref{eq:intro-E1+M1}) to include up to the electric $2^{n+1}$-pole and the magnetic $2^m$-pole interactions, referred to as ${\rm LG} (n,m)$ hereafter [more rigorous expressions of ${\rm LG} (n,m)$ are given below in Eqs.~(\ref{eq:LGnm}) and (\ref{HamEleGen}). 
A nonlinear term is to be added to Eq.~(\ref{eq:intro-E1+M1})]. 
Similarly, we can further extend Eq.~(\ref{eq:intro-VG1}) to expand ${\bf A}({\bf r},t)$ to the $\ell$-th order in ${\bf r}$ [${\rm VG} (\ell)$] and Eq.~(\ref{eq:intro-VG1prime}) to expand $[{\bf \hat{p}}+\A({\bf r},t)]^2$ to the $\ell$-th order in ${\bf r}$ [${\rm VG}^\prime (\ell)$].
Alternative gauges have also been proposed recently \cite{Forre2016PRA,Kjellsson2017PRA2}.

Aside from numerical efficiency \cite{Forre2014PRA,Forre2016PRA,Kjellsson2017PRA,Kjellsson2017PRA2}, a fundamental question is ``Are Eqs.~(\ref{eq:intro-E1+M1}) and (\ref{eq:intro-VG1}) [or (\ref{eq:intro-VG1prime})] mutually related through gauge transformation, so that we can obtain gauge-independent values of observable quantities? 
More generally, which ${\rm LG} (n,m)$ and ${\rm VG} (\ell)$ (or ${\rm VG}^\prime (\ell)$) are gauge-transformed to each other, and by what form of unitary operator?"
Such information will be useful, e.g., when we compare the simulation results from different numerical implementations.
We naively thought that the problems had already been investigated and searched for literature.
Although there are many works on gauge invariance and transformation (see Refs.~\cite{fiutak1963,kielich1965,barron1973,selsto2007} as well as a tutorial by Bandrauk {\it et al.}~\cite{bandrauk2013} and references therein), we could find very few papers possibly relevant to our questions.
Fiutak \cite{fiutak1963} transformed the exact VG Hamiltonian $H_{\rm VG}$ [see Eq.~(\ref{HamVelInf}) below] by a truncated unitary operator [$W_1$ in our notation Eq.~(\ref{GaugeGenk})] and obtained an LG-like Hamiltonian with the electric dipole, magnetic dipole, and electric quadrupole interactions.
It also contained, however, terms involving higher derivatives of ${\bf A}$ whose precise forms were not specified. 
While Selst{\o} and F{\o}rre \cite{selsto2007} have proposed transformation via $\e^{- \im \rr\cdot\A(\rr,t)}$, the resulting Hamiltonian does not explicitly contain the magnetic dipole, electric quadrupole, $\cdots$ terms.

In this paper, we study the gauge correspondence between ${\rm VG} (\ell)$ and ${\rm LG} (n,m)$. 
First, in Sec.~\ref{sec:Multipole Expansion}, we revisit the derivation of multipole expansion from the minimal coupling Hamiltonian and 
present a generic tensor-algebraic expression for the Hamiltonian corresponding to ${\rm LG} (n,m)$ for an arbitrary pair of nonnegative integers $n$ and $m$.
It contains not only electric and magnetic multipoles but also two additional terms.
Then, in Sec.~\ref{sec:Gauge Transformation}, we show that there exists gauge invariance between ${\rm VG} (\ell)$ and ${\rm LG} (n,m)$ if $\ell = n = m$, and give the form of the bridging unitary operator.
The additional terms are indispensable for the equivalence.
On the other hand, we could not find transformation between ${\rm VG}^\prime (\ell)$ and ${\rm LG} (n,m)$.
 
\section{Multipole Expansion}
\label{sec:Multipole Expansion}

Let us consider a single particle with mass $M$ and charge $q$.
The external field is assumed to be purely classical and not affected by the particle. 
In our notation, the operator $\nabla$ acts on the particle coordinate, denoted by $\rr$.
We apply the Coulomb gauge condition $\del_{\xx}\cdot\A(\xx,t)=0$ \cite{loudon2000,cohen2004}, where the external electromagnetic wave is transverse and described by the vector potential $\A(\xx,t)$.
The laser electric field is given by $\E(\xx,t) = -\dot{\A}(\xx,t)$, and the magnetic field by $\B(\xx,t) = \del_{\xx}\times\A(\xx,t)$.
Also, for simplicity, we drop the scalar potential that typically describes the Coulomb force from other classically treated charges.
It is straightforward to extend our discussion to a system with many particles and a scalar potential, since the interparticle Coulomb interaction and the scalar potential do not change their forms under gauge transformation.

We start from the minimal coupling Hamiltonian,
\beq\label{HamVelInf}
H_{\rm VG} = \frac{[- \im \nabla - q\A(\rr,t)]^2}{2M},
\eeq
considered to be the exact velocity gauge Hamiltonian of infinite order. 
The exact length-gauge ($\E$-$\B$) Hamiltonian is obtained through the Power-Zienau-Woolley transformation \cite{fiutak1963, andrews2018},
\beq\label{UnitTrans}
{H}_{\rm LG} = W^{-1}{H}_{\rm VG}W - \im W^{-1}\dot{W},
\eeq
where the unitary operator $W$ is defined as,
\beq\label{GaugeGen}
W = \e^{\im q\chi(\rr,t)},\quad \chi(\rr,t) = \int_0^1\rr\cdot\A(\lambda\rr,t)\de\lambda.
\eeq
%
The resultant Hamiltonian is given by \cite{loudon2000,cohen2004},
\beq\nonumber
\label{HamLenInf}
H_{\rm LG} &=& \frac{1}{2M}\left[ - \im \nabla + q\int_0^1\lambda\rr\times \B(\lambda\rr,t)\de\lambda  \right]^2 \\&-& q\int_0^1\rr\cdot\E(\lambda\rr,t)\de\lambda.
\eeq
The VG wave function $\psi_{\rm VG}({\bf r},t)$ is transformed to the LG one $\psi_{\rm LG}({\bf r},t)$ as $\psi_{\rm LG}({\bf r},t)=W^{-1}\psi_{\rm VG}({\bf r},t)$.

In what follows, we omit $t$ whenever clearly understood. By truncating the expansion of $\A(\xx)$ in Eq.~(\ref{HamVelInf}) at the $\ell$-th order of $\xx$, we obtain,
\beq\label{HamVelGen}
H_{\rm VG}^{(\ell)} = \frac{[ - \im \nabla - q\A^{(\ell)}(\rr)]^2}{2M},
\eeq
referred to as ${\rm VG}(\ell)$, with
\begin{equation}
	\label{eq:Atruncated}
	\A^{(\ell)}(\rr) = \sum_{k=0}^\ell\frac{(\rr\cdot\del_\xx)^k\A(\xx)|_{\xx=0}}{k!}.
\end{equation}
${\rm VG}(0)$ corresponds to EDA Eq.~(\ref{eq:intro-VG0}). We call a term involving $q^2$, such as $q^2|\A^{(\ell)}(\rr)|^2/(2M)$, a {\it nonlinear term}.

Similarly, let us truncate $\E(\xx)$ and $\B(\xx)$ in Eq.~(\ref{HamLenInf}) at the $n$-th and $(m-1)$-th order, respectively, of $\xx$, and call the resulting gauge ${\rm LG}(n,m)$:
\beq\nonumber
H_{\rm LG}^{(n,m)} &=& \frac{1}{2M}\left[ - \im \nabla + q\int_0^1\lambda\rr\times \B^{(m-1)}(\lambda\rr)\de\lambda  \right]^2 \\&-& q\int_0^1\rr\cdot\E^{(n)}(\lambda\rr)\de\lambda,
\label{eq:LGnm}
\eeq
where $\E^{(\ell)}$ and $\B^{(\ell)}$ are defined in a way similar to $\A^{(\ell)}$. $\B^{(-1)}$ is taken to be zero. 
It follows from the definition of each gauge that the relation between the canonical momentum $\hat{\bf p} = -\im\nabla$ and the kinetic momentum $\hat{\bf \pi}$ is given by,
\begin{equation}
	\hat{{\bf\pi}} = \left\{
	\begin{array}{ll}
	\hat{\bf p} - q\A(\rr) & {\rm VG} \\ 
	\hat{\bf p} - q\A^{(\ell)}(\rr) & {\rm VG}(\ell) \\ 
	\hat{\bf p} + q\int_0^1\lambda\rr\times\B(\lambda\rr)\de\lambda & {\rm LG} \\ 
	\hat{\bf p} + q\int_0^1\lambda\rr\times\B^{(m-1)}(\lambda\rr)\de\lambda & {\rm LG}(n,m)
	\end{array}\right. .
\end{equation}
Thus, $\hat{\bf p} \neq \hat{\bf \pi}$ except for in the case of LG($n,0$). 
This is related to the fact that $-\int_0^1\lambda\rr\times\B(\lambda\rr)\de\lambda$ and $-\int_0^1\rr\cdot\E(\lambda\rr)\de\lambda$ are the vector and scalar potentials, respectively, in the Poincar\'e gauge \cite{cohen2004}.

We obtain multipole expansion by carrying out the integration with respect to $\lambda$ in Eq.~(\ref{eq:LGnm}).
With the electric dipole $\mathbf{d} = q\rr$ and the magnetic dipole $\mathbf{m} = -\im q\rr\times\nabla/M$, the first few examples are given by,
\beq
{H_{\rm LG}^{(0,0)}} = -\frac{\nabla^2}{2M} - \mathbf{d}\cdot\E(0),
\eeq
\beq\label{LGHamM1}\nonumber
{H_{\rm LG}^{(0,1)}} = -\frac{\nabla^2}{2M} - \mathbf{d}\cdot\E(0) - \frac{1}{2}\mathbf{m}\cdot\mathbf{B}(0) +\frac{q^2(\rr\times\mathbf{B}(0))^2}{8M},\\
\eeq
\beq\label{LGHamE2}\nonumber
{H_{\rm LG}^{(1,0)}} = -\frac{\nabla^2}{2M} - \mathbf{d}\cdot\E(0) - \frac{1}{2}\sum_{ij}qr^ir^j\frac{\del E_j(\xx)}{\del x^i}\bigg|_{\xx=0},\\
\eeq
\beq\label{LGHamE2full}\nonumber
{H_{\rm LG}^{(1,1)}} &&= -\frac{\nabla^2}{2M} - \mathbf{d}\cdot\E(0) - \frac{1}{2}\mathbf{m}\cdot\mathbf{B}(0)\\ &&  - \frac{1}{2}\sum_{ij}qr^ir^j\frac{\del E_j(\xx)}{\del x^i}\bigg|_{\xx=0} +\frac{q^2(\rr\times\mathbf{B}(0))^2}{8M}.
\eeq
${H_{\rm LG}^{(0,0)}}$ corresponds to EDM Eq.~(\ref{eq:intro-LG0}). 
While the third term of Eq.~(\ref{LGHamM1}) is equivalent to that of Eq.~(\ref{eq:intro-E1+M1}), it should be noticed that Eq.~(\ref{LGHamM1}) contains the fourth, nonlinear term, which is absent in Eq.~(\ref{eq:intro-E1+M1}).

${H_{\rm LG}^{(1,0)}}$ and ${H_{\rm LG}^{(1,1)}}$ contains the electric quadrupole term $-\frac{1}{2}\sum_{ij}qr^ir^j\frac{\del E_j(\xx)}{\del x^i}\big|_{\xx=0}$.
Let us now rewrite the term using the tensor algebra (see Appendix \ref{AppTensor} for details). The {\it ad hoc} rule to perform the algebra with the tensor product $\otimes$ and vectors $\mathbf{v}_i\in\mathbb{R}^3$, dual vectors $\mathbf{u}_i^{\sf t}\in(\mathbb{R}^3)^*$ $(i=1,...,n)$ is, to interpret the contraction ``$:$" of two tensors consisting of the same number of vectors and dual vectors as 
\beq\label{TensorInnerProd}
\left[\bigotimes_{i=1}^n\mathbf{v}_i\right]:\left[\bigotimes_{j=1}^n\mathbf{u}^{\sf t}_j\right] = \prod_{k=1}^n(\mathbf{u}_k^{\sf t}\mathbf{v}_k),
\eeq
where $\bigotimes$ denotes the $n$-fold tensor product defined as,
\beq
\bigotimes_{i=1}^n\mathbf{v}_i = \overbrace{{\bf v}_1\otimes{\bf v}_2\otimes\cdots\otimes{\bf v}_{n-1}\otimes{\bf v}_n}^{n}.
\eeq
If we define the electric $2^n$-pole moment ${\mathcal Q}^{(n)}$ and the $n$-th electric gradient field ${\mathcal E}^{(n)}(\xx)$ by,
\beq
{\mathcal Q}^{(n)} &=& q\overbrace{{\rr}\otimes{\rr}\otimes\cdots\otimes{\rr}\otimes{\rr}}^{n} \equiv q\otimes^n\rr,\\ 
{\mathcal E}^{(n)}(\xx) &=& \otimes^{n-1}\del_{\xx}\otimes\E(\xx),
\eeq
and note that $\del_\xx$ and $\E$ are dual vectors, then, the electric quadrupole term can be rewritten as,
\beq
-\frac{1}{2}\sum_{ij}qr^ir^j\frac{\del E_j(\xx)}{\del x^i}\bigg|_{\xx=0} = - \frac{1}{2}{\mathcal Q}^{(2)}:{\mathcal E}^{(2)}(0).
\eeq

This tensor notation is fairly useful for the concise description of complicated higher terms. 
Let us also introduce the magnetic $2^n$-pole ${\mathcal M}^{(n)}$ and the $n$-th magnetic gradient field ${\mathcal B}^{(n)}(\xx)$ defined as,
\beq
{\mathcal M}^{(n)} &=& \otimes^{n-1}\rr\otimes{\bf m},\\ 
{\mathcal B}^{(n)}(\xx) &=& \otimes^{n-1}\del_\xx\otimes\B(\xx).
\eeq
Then, the general multipole-expansion form of the ${\rm LG}(n,m)$ Hamiltonian $H_{\rm LG}^{(n,m)}$ Eq.~(\ref{eq:LGnm}) for arbitrary $n$ and $m$ is expressed as,
\beq
\label{HamEleGen}\nonumber
{H}_{\rm LG}^{(n,m)} &=& -\frac{\nabla^2}{2M} - \sum_{k=1}^{{n+1}}\frac{1}{k!}{\mathcal Q}^{(k)}:{\mathcal E}^{(k)}(0) \\ \nonumber&&- \sum_{k=1}^{m}\frac{k}{(k+1)!}{\mathcal M}^{(k)}:{\mathcal B}^{(k)}(0)\\ 
\nonumber
&& +\frac{\im}{2Mc^2}\sum_{k=1}^{m-1}\frac{k(k+1)}{(k+2)!}{\mathcal Q}^{(k)}:\dot{\mathcal E}^{(k)}(0)
\\ \nonumber&& + \frac{q^2}{2M}\left[\rr\times\sum_{k=1}^m\frac{k}{(k+1)!}(\rr\cdot\del_\xx)^{k-1}\B(\xx)\big|_{\xx=0}\right]^2,\\
\eeq
where $\dot{\mathcal E}^{(k)}$ in the fourth term is introduced by using the Maxwell equation $\del_\xx\times\B(\xx) = \dot{\E}(\xx)/c^2$ with $c$ being the vacuum velocity of light.
We see that ${\rm LG}(n,m)$ for $n\ge 0$ and $m\ge 1$ contains all the multipoles up to the electric $2^{n+1}$-pole [E$(n+1)$] (second term) and the magnetic $2^m$-pole (M$m$) (third term) interactions, as expected.
It should, however, be noticed that Eq.~(\ref{HamEleGen}) contains, in addition, 
the fourth term for $m\ge 2$ and the fifth, nonlinear term for $m\ge 1$.
The limiting form of Eq.~(\ref{HamEleGen}) for $n,m\to\infty$ was given in Ref.~\cite{fiutak1963}.

The leading subterm of the fourth term is $-\frac{iq}{6Mc^2}\rr\cdot \dot{\E}(0)\approx -\frac{q\omega}{6Mc^2}\rr\cdot \E(0)$ with $\omega$ being the laser angular frequency.
For the electron ($q=-1,M=1$), its ratio to the electric dipole is $3.26\times 10^{-7}\hbar\omega/{\rm eV}$.
Similarly, the ratio of the leading subterm of the fifth term to the electric dipole is roughly $7\times 10^{-6}|\rr||\E|$.
Thus, these terms are in general much smaller than the electric dipole term and have usually been neglected. It is, however, not {\it a priori} clear if they are negligible compared with higher multipoles in a strong laser field.

 
\section{Gauge Transformation}
\label{sec:Gauge Transformation}

\begin{figure}
\centering
\includegraphics[width=\linewidth]{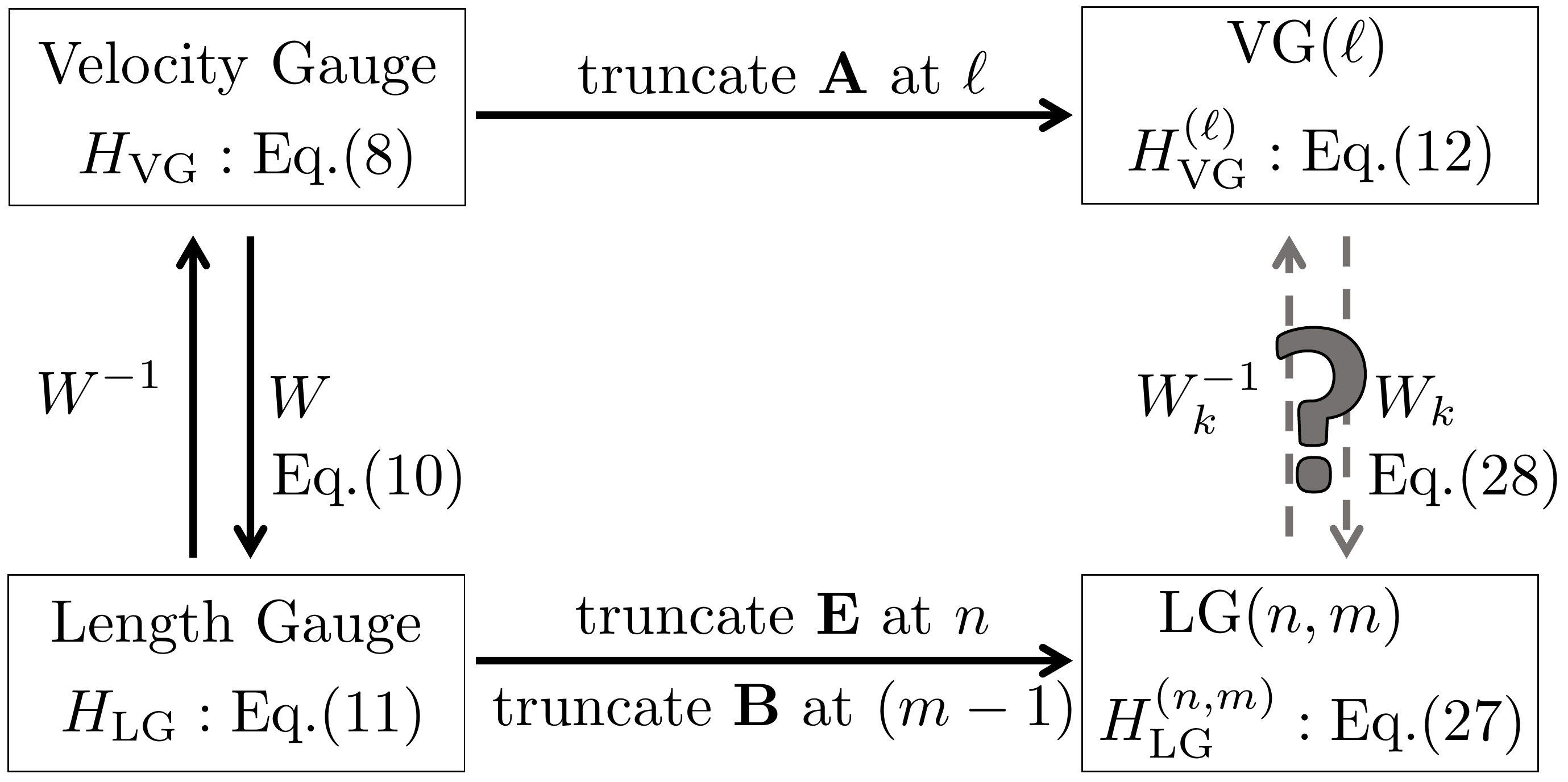}
\caption{Relation of the different gauges and Hamiltonians considered in this study.\label{fig:DiagCom}}
\end{figure}

Let us now proceed to the discussion on the gauge transformation.
We introduce a unitary operator,
\beq\label{GaugeGenk}
W_\ell (\rr) = \e^{iq \chi^{(\ell)}(\rr)},
\eeq
where,
\beq
\chi^{(\ell)}(\rr) &=& \int_0^1\rr\cdot\A^{(\ell)}(\lambda\rr)\de\lambda \\
&=& \rr\cdot\sum_{k=0}^\ell\frac{(\rr\cdot\del_{\xx})^k\A(\xx)\big|_{\xx=0}}{(k+1)!}.
\eeq
%
%
%
Note that $\chi^{(0)}({\bf r}) = {\bf r}\cdot {\bf A}(0)$ and that $\chi^{(\ell)}({\bf r})$ is the expansion of $\chi ({\bf r})$ to the $(\ell+1)$-th order with respect to ${\bf r}$.
Then, we transform the ${\rm VG}(\ell)$ Hamiltonian via $W_k$ to,
\beq
{H}_{\rm VG\to}^{(\ell|k)} = W_k^{-1}{H}_{\rm VG}^{(\ell)}W_k - \im W^{-1}_k\dot{W}_k,
\eeq
and the ${\rm LG}(n,m)$ Hamiltonian via $W_k^{-1}$ to,
\beq
{H}_{\rm LG\to}^{(n,m|k)} = W_k{H}_{\rm LG}^{(n,m)}W_k^{-1} - \im W_k\dot{W}_k^{-1}.
\eeq
In Fig.~\ref{fig:DiagCom} we summarize the relation and transformation between different gauges considered in this study.
It may be worth mentioning at this point that Fiutak \cite{fiutak1963} considered ${H}_{\rm VG\to}^{(\infty|1)}$ and that Bandrauk {\it et al.}~\cite{bandrauk2013} proposed ${H}_{\rm VG\to}^{(1|0)}$.

We list low-order examples:
\beq
\label{eq:gauge-transform-000}
{H}_{{\rm LG}\to}^{(0,0|0)} = {H_{\rm VG}^{(0)}}\quad,\quad {H}_{{\rm VG}\to}^{(0|0)}  = {H_{\rm LG}^{(0,0)}},
\eeq
\beq
\nonumber
{H}_{{\rm LG}\to}^{(0,1|0)} &=&  \frac{[ - \im \nabla - q\A(0)]^2}{2M} - \frac{1}{2}\mathbf{m}\cdot\mathbf{B}(0) \\
&+& \frac{q^2\A(0)\cdot[\rr\times\B(0)]}{2M} + \frac{q^2(\rr\times\mathbf{B}(0))^2}{8M}, 
\label{eq:gauge-transform-010}
\eeq
\beq\label{eq:gauge-transform-111}
{H_{{\rm LG}\to}^{(1,1|1)}} = {H_{\rm VG}^{(1)}}\quad,\quad {H}_{{\rm VG}\to}^{(1|1)} = {H_{\rm LG}^{(1,1)}}\quad,
\eeq
\beq
\label{eq:gauge-transform-10}
{H}_{{\rm VG}\to}^{(1|0)} = {H_{\rm LG}^{(0,0)}} + \frac{\im q(\rr\cdot\del_{\xx})\A(\xx)\big|_{\xx=0}}{M}\cdot\nabla \quad,
\eeq
The comparison between Eqs.~(\ref{LGHamM1}) and (\ref{eq:gauge-transform-010}) shows that, while the electric dipole term is transformed, the magnetic dipole term $- \frac{1}{2}\mathbf{m}\cdot\mathbf{B}(0)$ remains unchanged, and Eq.~(\ref{eq:gauge-transform-010}) contains an additional nonlinear term (third term).
Thus, $W_0$ does not map Eq.~(\ref{eq:intro-E1+M1}) to a Hamiltonian of a velocity-gauge form.


Equations~(\ref{eq:gauge-transform-000}) and (\ref{eq:gauge-transform-111}) suggest that $W_n$ transforms $H_{\rm VG}^{(n)}$ to $H_{\rm LG}^{(n,n)}$, and vice versa, i.e., 
\begin{equation}
	H_{\rm VG\to}^{(n|n)} = H_{\rm LG}^{(n,n)} \quad , \quad H_{\rm LG\to}^{(n,n|n)} = H_{\rm VG}^{(n)}.
\end{equation}
We can demonstrate that this indeed holds true for any $n(\ge 0)$ as follows.
$\E^{(n)}$ and $\B^{(n-1)}$ are related with $\A^{(n)}$ by $\E^{(n)}(\xx,t) = -\dot{\A}^{(n)}(\xx,t)$ and $\B^{(n-1)}(\xx,t) = \del_{\xx}\times\A^{(n)}(\xx,t)$, respectively.
These are analogous to the relations of $\E$, $\B$, and $\A$. Then, in the same way as the derivation of Eq.~(\ref{HamLenInf}) from Eq.~(\ref{HamVelInf}), one can show that $H_{\rm VG}^{(n)}$ is transformed to $H_{\rm LG}^{(n,n)}$ [see Eq.~(\ref{eq:LGnm})] via $W_n$.
We also present an explicit derivation in Appendix \ref{AppDer}.
The ${\rm VG}(n)$ and ${\rm LG}(n,n)$ wave functions $\psi_{\rm VG}^{(n)}({\bf r},t)$ and $\psi_{\rm LG}^{(n,n)}({\bf r},t)$, respectively, are related to each other via $\psi_{\rm VG}^{(n)}=W_n\psi_{\rm LG}^{(n,n)}$.


Thus, ${\rm VG}(n)$ and ${\rm LG}(n,n)$ are physically equivalent, i.e., numerically exact simulations with ${H}_{\rm VG}^{(n)}$ and ${H}_{\rm LG}^{(n,n)}$ would yield the same value for any observable.
This is reasonable, since $\E^{(n)}$ and $\B^{(n-1)}$ are the electric and magnetic fields, respectively, described by the vector potential $\A^{(n)}$ ; in the classical electrodynamics, Hamilton's equations of motion with $H_{\rm VG}^{(n)}$ lead to Newton's for a charged particle under $\E^{(n)}$ and $\B^{(n-1)}$ (see Appendices \ref{sec:Eq6} and \ref{sec:Eq7}).
It should, however, be stressed that we should retain the fourth and fifth terms of Eq.~(\ref{HamEleGen}) and the nonlinear term $q^2|\A^{(n)}(\rr)|^2/(2M)$ in Eq.~(\ref{HamVelGen}), from the conceptual viewpoint of the gauge invariance \cite{selsto2007}.
On the other hand, if $n^\prime\ne n$ and/or $m^\prime\ne n$, ${H}_{\rm LG}^{(n^\prime,m^\prime)}$ and ${H}_{\rm LG}^{(n,n)}$, with different multipoles, obviously correspond to distinct physical situations and, thus, would lead to different observable values.
Therefore, ${H}_{\rm LG}^{(n,m)}$ and ${H}_{\rm VG}^{(\ell)}$ can be transformed to each other if and only if $n=m=\ell$ (see Appendix \ref{AppPro} for another proof).

\section{Conclusions}
\label{sec:Conclusions}

We have investigated correspondence between the velocity and length gauges of laser interaction with charged particles, beyond the electric dipole approximation.
After presenting the length-gauge Hamiltonian for arbitrary orders of multipole expansion, we have shown that ${H}_{\rm VG}^{(\ell)}$ [Eq.~(\ref{HamVelGen})] and ${H}_{\rm LG}^{(n,m)}$ [Eq.~(\ref{HamEleGen})] can be mutually gauge-transformed if $n=m=\ell$, via $W_n$ [Eq.~(\ref{GaugeGenk})].
The fourth and nonlinear fifth terms in Eq.~(\ref{HamEleGen}) and the nonlinear term in Eq.~(\ref{HamVelGen}) should be included to ensure exact physical equivalence.
It may come as a surprise that this seemingly fundamental issue has not been explicitly addressed before.
It is probably partially because light-matter interaction has conventionally been studied by use of truncated perturbation expansion with respect to field strength, where gauge invariance is anyway lost.
In strong-field and attosecond physics, in contrast, we often encounter extremely nonlinear processes, which has driven activity to develop nonperturbative time-dependent methods \cite{Kulander1992} (see also Ref.~\cite{Ishikawa2015JSTQE} and references therein).
Then, the gauge invariance (or dependence) of numerical approaches has become an important issue.
Hence, the present finding will be beneficial to numerical study of phenomena that require treatments beyond EDA, such as the response of chiral molecules and nanostructured materials \cite{Iwasa2009PRA}. 

\begin{acknowledgments}

This research was supported in part by Grants-in-Aid for Scientific Research (Grants No.~16H03881, No.~17K05070, No.~18K14145, and 18H03891) and by Exploratory Challenge on Post-K Computer from the Ministry of Education, Culture, Sports, Science and Technology (MEXT) of Japan and also by the Photon Frontier Network Program of MEXT. This research was also partially supported by the Center of Innovation Program from the Japan Science and Technology Agency, JST, and by CREST (Grant No. JPMJCR15N1), JST. R.A. gratefully acknowledges support from the Graduate School of Engineering, The University of Tokyo, Doctoral Student Special Incentives Program (SEUT Fellowship).

\end{acknowledgments}

\appendix

\section{Tensor Algebra on Euclidean Space}\label{AppTensor}
Let us assume that we have an $n$-dimensional real metric vector space whose metric tensor is the unit tensor (i.e., $n$-dimensional Euclidean space), denoted by $\mathbb{R}^n$.  We also denote its natural orthonormal basis by $\{\mathbf{e}_1,...,\mathbf{e}_n\}$ and its dual basis by $\{\mathbf{e}^{\sf t}_1,...,\mathbf{e}^{\sf t}_n\}$, such that 
\beq
{\bf e}_{i} = \left[\begin{array}{c}\delta_{1i}\\ \delta_{2i}\\\vdots\\ \delta_{ni} \end{array}\right]; \quad {\bf e}^{\sf t}_j = \left[\delta_{1j}, \delta_{2j},..., \delta_{nj}\right],
\eeq
with the unit tensor $\delta_{ii} = 1$ and $\delta_{ij} = 0$ for any $i\neq j$. 
Let us denote an arbitrary element of $\mathbb{R}^n$ by ${\bf v}$, and its dual (transposed) vector by ${\bf v}^{\sf t}\in(\mathbb{R}^n)^*$. Note that the inner product is $({\bf v},{\bf u}) = ({\bf u}, \bf{v})= {\bf u}^{\sf t}{\bf v}$, where in the last term the usual matrix multiplication is used. Then, a tensor ${\sf T}$ of type $(a,b)$ is defined as a multilinear function
\beq
{\sf T}: \overbrace{(\mathbb{R}^n)^*\otimes\cdots\otimes (\mathbb{R}^n)^*}^{a}\otimes\overbrace{\mathbb{R}^n\otimes\cdots\otimes \mathbb{R}^n}^{b}
\to 
\mathbb{R}.
\eeq
Here $\otimes$ is the direct product of vector spaces. The vector space consisting of type $(a,b)$ tensors is called the {\it tensor space} of kind $(a,b)$, denoted by $T^a_b(\mathbb{R}^n)$. Note that $T^0_0(\mathbb{R}^n) =\mathbb{R}$, $T^1_0(\mathbb{R}^n) =\mathbb{R}^n$, and $T^0_1(\mathbb{R}^n) = (\mathbb{R}^n)^*$. 
For any ${\sf T} \in T_b^a(\mathbb{R}^n)$, there is a unique representation $T^{i_1\cdots i_a}_{j_1\cdots j_b}$ such that
\beq
{\sf T} = \sum_{i_1\cdots i_a}\sum_{j_1\cdots j_b}T_{j_1\cdots j_b}^{i_1\cdots i_a}\bigotimes_{k=1}^a{\bf e}_{i_k}\bigotimes_{\ell=1}^b{\bf e}^{\sf t}_{j_\ell}.
\eeq
For two tensors ${\sf T}\in T^a_0(\mathbb{R}^n)$ and ${\sf S}\in T_a^0(\mathbb{R}^n)$, we define the contraction "$:$" as
\beq
{\sf T}:{\sf S} = \sum_{i_1\cdots i_a}T^{i_1\cdots i_a}S_{i_1\cdots i_a},
\eeq
which means, in a special case relevant in our context,
\beq
\left[\bigotimes_{i=1}^n\mathbf{v}_i\right]:\left[\bigotimes_{j=1}^n\mathbf{u}^{\sf t}_j\right] = \prod_{k=1}^n(\mathbf{u}_k^{\sf t}\mathbf{v}_k).
\eeq
This is equivalent to Eq.~(\ref{TensorInnerProd}).

\section{Explicit Derivation of Equivalence between $H_{\rm LG}^{(n,n)}$ and $H_{\rm VG}^{(n)}$}
\label{AppDer}

For the sake of simple notation, we introduce,
\beq\label{DefA}
	\A^{[\ell]}(\xx) &=& (\xx\cdot\del_{\xx'})^\ell\A(\xx')\big|_{\xx'=0}\quad, \\
	{\underline{\A}}^{[\ell]}(\xx) &=& \del_{\xx'}(\xx\cdot\del_{\xx'})^{\ell-1}(\xx\cdot\A(\xx'))\big|_{\xx'=0}\quad,\\
	\tilde{\A}^{(\ell)}(\xx) &=& \sum_{k=0}^\ell\frac{\A^{[k]}(\xx)}{(k+1)!}, 
\eeq
and a unitary operator, 
\beq\label{GaugeGenk}
W_\ell (\rr) = \e^{iq \chi^{(\ell)}(\rr)},\quad \chi^{(\ell)}(\rr) = \rr\cdot\tilde{\A}^{(\ell)}(\rr).
\eeq
Then the differential operator $\nabla$ transforms according to $W_n$ as
\beq\nonumber
W_n^{-1}\nabla W_n &=& \nabla +\im q\nabla\chi^{(n)}\\ &=& \nabla + \im q\left[\tilde{\bf A}^{(n)} + \sum_{k=0}^n\frac{k{\underline{\A}}^{[k]}}{(k+1)!}\right],
\eeq
that the transformed covariant derivative reads,
\beq\nonumber\label{CovDer}
	&& W_n^{-1}[- \im \nabla - q\A^{(n)}]W_n\\ 
	\nonumber && = -\im \nabla + q\left[-\A^{(n)}+\tilde{{\A}}^{(n)} + \sum_{k=0}^n\frac{k{\underline{\A}}^{[k]}}{(k+1)!}\right] \\
	&& =-\im \nabla + q\sum_{k=0}^n\frac{k\left({\underline{\A}}^{[k]}-\A^{[k]}\right)}{(k+1)!}.
\eeq
Noting that
\beq\label{XcrossB}
{\underline{\A}}^{[k]}(\xx) - \A^{[k]}(\xx) = \xx\times(\xx\cdot\del_{\xx'})^{k-1}\B(\xx')\big|_{\xx'=0},
\eeq
we obtain,
\beq\nonumber
&& W_n^{-1}H_{\rm VG}^{(n)}W_n\\
&& \quad = \frac{1}{2M}\left[-\im \nabla + q\int_0^1\lambda\rr\times \B^{(n-1)}(\lambda\rr)\de\lambda\right]^2.
\eeq
Hence, the transformed Hamiltonian is given by,
\beq\nonumber\label{EQIV}
H_{{\rm VG}\to}^{(n|n)} &=&  \frac{1}{2M}\left[-\im \nabla + q\int_0^1\lambda\rr\times \B^{(n-1)}(\lambda\rr)\de\lambda\right]^2\\ &&+ q\int_0^1\rr\cdot\dot\A^{(n)}(\lambda\rr)\de\lambda \\
&=& H_{\rm LG}^{(n,n)}.
\eeq
This also means that $H_{{\rm LG}\to}^{(n,n|n)} = H_{{\rm VG}}^{(n)}$.

\section{Newtonian equation of motion derived from VG($\ell$)}
\label{sec:Eq6}

Using the VG($\ell$) Hamiltonian [Eq.(\ref{HamVelGen})],
\beq
H_{\rm VG}^{(\ell)} = \frac{[\PP - q\A^{(\ell)}]^2}{2M},
\eeq
we start from the classical canonical equations (we put $\PP = (p_x,p_y,p_z)^{\rm T}$),
\beq\label{ceq}
\frac{\de r_i}{\de t} = \frac{H_{\rm VG}^{(\ell)}}{\del p_i}, \quad \frac{\de {p}_i}{\de t} = -\frac{H_{\rm VG}^{(\ell)}}{\del r_i}.
\eeq
The first equation leads to,
\beq
\frac{\de r_i}{\de t}  = \frac{p_i - qA_i^{(\ell)}}{M}.
\eeq
Therefore, the velocity ${\bf v} = \frac{\de\rr}{\de t}$ is given by,
\beq\label{vel}
{\bf v} = \frac{\PP - q\A^{(\ell)}}{M}.
\eeq
The second equation in Eq.(\ref{ceq}) becomes, for the $x$-component, 
\beq\nonumber\label{mom}
\frac{\de p_x}{\de t} &=& -\frac{\del H_{\rm VG}^{(\ell)}}{\del x}\\ \nonumber
&=& -\frac{\PP - q\A^{(\ell)}}{M}\cdot\frac{-q\del\A^{(\ell)}}{\del x}\\ 
&=& q\frac{\PP - q\A^{(\ell)}}{M}\cdot\frac{\del\A^{(\ell)}}{\del x}.
\eeq
Then, the $x$-component of the Newtonian equation of motion is given by, using the derivative $\nabla$ with respect to $\xx$,
%
%
\beq\nonumber
&&M\frac{\de v_x}{\de t} = \frac{\de p_x}{\de t} - q\frac{\de A_x^{(\ell)}}{\de t}\\ && \quad = q{\bf v}\cdot\frac{\del\A^{(\ell)}}{\del x} - q({\bf v}\cdot\nabla)A_x^{(\ell)} - q\frac{\del A_x^{(\ell)}}{\del t}.
\eeq
%
Thus, using the vector formula $\A\times(\B\times\C) = (\A\cdot\C)\B-(\A\cdot\B)\C$ and noting $\B^{(\ell-1)} = \nabla\times\A^{(\ell)}$, we obtain,
\beq\label{LF}
M\frac{\de {\bf v}}{\de t} = q\E^{(\ell)} + q{\bf v}\times\B^{(\ell-1)}.
\eeq
The right-hand side is the physically consistent and correct expression for the Lorentz force within the considered truncation $\E^{(\ell)}$ and $\B^{(\ell-1)}$.

\section{Newtonian equation of motion derived from ${\rm VG}^\prime$(1)}
\label{sec:Eq7}

Using the ${\rm VG}^\prime (1)$ Hamiltonian [Eq.(7)],
\beq\nonumber
	&&H_{{\rm VG}^\prime}^{\rm (1)}= \frac{[{\PP}-q\A(0)]^2}{2M} 
	\\  &&\quad - \frac{q}{M}(\rr\cdot\nabla)\A(\xx,t)\big|_{\xx=0}\cdot[{\PP}-q\A(0)],
\eeq
we start from the classical canonical equations,
\beq\label{ceq7}
\frac{\de r_i}{\de t} = \frac{H_{\rm VG'}^{(1)}}{\del p_i}; \quad \frac{\de {p}_i}{\de t} = -\frac{H_{\rm VG'}^{(1)}}{\del r_i}.
\eeq
The first equation leads to,
\beq
\frac{\de r_i}{\de t} = \frac{p_i - qA_i(0) - q(\rr\cdot\nabla)A_i(\xx,t)\big|_{\xx=0}}{M}.
\eeq
Therefore, the velocity ${\bf v} = \frac{\de \rr}{\de t}$ is given by,
\beq\nonumber
{\bf v} &=& \frac{\PP - q\A(0) - q(\rr\cdot\nabla)\A(\xx,t)\big|_{\xx=0}}{M}\\ &=& \frac{\PP - q\A^{(1)}}{M} 
\eeq
The second equation in Eq.(\ref{ceq7}) becomes, for the $x$-component,
\beq\nonumber\label{mom7}
\frac{\de p_x}{\de t} &=& -\frac{\del H_{\rm VG'}^{(1)}}{\del x}\\ \nonumber
&=& -\frac{\PP - q\A(0)}{M}\cdot\frac{-q\del\A(\xx)}{\del x}\bigg|_{\xx=0}\\ 
&=& q\frac{\PP - q\A(0)}{M}\cdot\frac{\del\A^{(1)}(\xx)}{\del x}\bigg|_{\xx=0}.
\eeq
Then, the $x$-component of the Newtonian equation of motion is given by,
\begin{widetext}
\beq\nonumber
 M\frac{\de^2 x}{\de t^2} &=& M\frac{\de v_x}{\de t} = \frac{\de p_x}{\de t} - q\frac{\de A_x^{(1)}}{\de t}\\ \nonumber &=& \frac{\de p_x}{\de t} - q({\bf v}\cdot\nabla)A_x^{(1)}\big|_{\xx=0} - q\frac{\del A_x^{(1)}}{\del t} \\
&=& qE_x^{(1)} + q\left[\left({\bf v} + \frac{q(\rr\cdot\nabla)\A(\xx,t)\big|_{\xx=0}}{M}\right)\cdot\frac{\del\A^{(1)}}{\del x}\Bigg|_{\xx=0}- ({\bf v}\cdot\nabla)A_x^{(1)}\bigg|_{\xx=0}\right].
\eeq
Using the vector formula $\A\times(\B\times\C) = (\A\cdot\C)\B-(\A\cdot\B)\C$, we otbain,
\beq
\label{eq:NEOM7}
M\frac{\de {\bf v}}{\de t} = q\E^{(1)} + q{\bf v}\times\B^{(0)} + \frac{q^2(\rr\cdot\nabla)\A(\xx,t)\big|_{\xx=0}\cdot\del\A^{(1)}(\xx')}{M\del\xx'}\Bigg|_{\xx'=0}
\eeq
\end{widetext}
This equation contains not only the Lorentz force but also an extra (third) term, which is difficult to interpret physically.

\section{Another proof of non-equivalence between VG$(\ell)$ and LG$(n,m)$ for $(n,m)\neq(\ell,\ell)$}
\label{AppPro}

We seek a unitary operator $V$ that transforms VG$(\ell)$ to LG$(n,m)$, reading
\beq\label{Cond1}
H_{\rm LG}^{(n,m)} = V^{-1}H_{\rm VG}^{(\ell)}V - iV^{-1}\dot{V}.
\eeq
We show that if we suppose $(n,m)\neq(\ell,\ell)$ above, it will lead to a contradiction (for the sake of {\it reductio ad absurdum}). The unitary operator $V$ can be expressed as the product $V = W_\ell U$. The equation for $V$ now reads
\beq
H_{\rm LG}^{(n,m)} = U^{-1}H_{\rm LG}^{(\ell,\ell)}U - iU^{-1}\dot{U}.
\eeq
We introduce a real-valued function $u = u(\rr,t)$ such that $U = \e^{iu}$. Then the equation above becomes ($\pp = -i\nabla$),
\beq\nonumber
&&H_{\rm LG}^{(n,m)}\\ \nonumber &&= \frac{1}{2M}\left[\pp + q\int_0^1\de\lambda~\lambda\rr\times\B^{(\ell-1)}(\lambda\rr,t) + \nabla u(\rr,t)\right]^2\\ &&- q\int_0^1\de\lambda\rr\cdot\E^{(\ell)}(\lambda\rr,t) + \dot{u}(\rr,t).
\eeq
Since the mass $M$ is independent from this transformation, we can compare the mass-free terms in both sides, yielding
\beq\label{dotu}\nonumber
\dot{u}(\rr,t) = -q\int_0^1\de\lambda\rr\cdot\left[ \E^{(n)}(\lambda\rr,t) - \E^{(\ell)}(\lambda\rr,t)\right].\\
\eeq
Also, comparing coefficients for $\pp$ in both sides, we obtain
\beq\label{gradu}\nonumber
\nabla u(\rr,t) = q\int_0^1\de\lambda\rr\times\left[\B^{(m-1)}(\lambda\rr,t) - \B^{(\ell-1)}(\lambda\rr,t)\right].\\
\eeq
The function $u = u(\rr,t)$ must satisfy these equations simultaneously. 

First we assume $n=\ell$. Form Eq.(\ref{dotu}), it is concluded that $\dot{u} = 0$ in this case. However, if this is true, Eq.(\ref{gradu}) cannot hold, since the right-hand side is explicitly time-dependent while $\nabla u$ has no time-dependencies.

Next we assume $n>\ell$. In this case, from Eq.(\ref{dotu}) the function $u = u(\rr,t)$ is a polynomial of order $n+1$, starting from order $\ell+2$. Thus the function $\nabla u(\rr,t)$ is a polynomial of order $\ell+1$ to $n$. Hence by Eq.(\ref{gradu}), the solution for Eqs.(\ref{dotu}) and (\ref{gradu}) exists only if $n=m$. This is also true for $n<\ell$. Taking the temporal integral for $\dot{u}$, we have
\beq\nonumber
u(\rr,t)= q\int_0^1\de\lambda\rr\cdot\left[\A^{(n)}(\lambda\rr,t) - \A^{(\ell)}(\lambda\rr,t)\right] + u_0(\rr),\\
\eeq
with $u_0$ being the integral constant. Taking the gradient of above, we obtain 
\beq\label{newgradu}\nonumber
&&\nabla u(\rr,t)\\ \nonumber &&= q\nabla\int_0^1\de\lambda\rr\cdot\left[\A^{(n)}(\lambda\rr,t) - \A^{(\ell)}(\lambda\rr,t)\right] + \nabla u_0(\rr).\\
\eeq 
By substituting $\nabla u$ in Eq.(\ref{gradu}) by the right-hand side of Eq.(\ref{newgradu}), we obtain
\beq\nonumber
&&q\int_0^1\de\lambda\rr\times\left[\B^{(n-1)}(\lambda\rr,t) - \B^{(\ell-1)}(\lambda\rr,t)\right] \\ \nonumber &&= q\nabla\int_0^1\de\lambda\rr\cdot\left[\A^{(n)}(\lambda\rr,t) - \A^{(\ell)}(\lambda\rr,t)\right] + \nabla u_0(\rr).\\
\eeq
By means of Eq.(\ref{XcrossB}), the equation above yields
\beq
q\left[\A^{(\ell)}(\rr,t) - \A^{(n)}(\rr,t)\right] = \nabla u_0(\rr).
\eeq
This equation leads to a contradiction, since the left-hand side is explicitly time-dependent, while the right-hand side is a constant in time.

Hence we have contradictions in both cases ($n=\ell$ or $n\neq\ell$), provided $(n,m)\neq(\ell,\ell)$ in Eq.(\ref{Cond1}). Therefore there are no unitary operators that satisfies Eq.(\ref{Cond1}) for $(n,m)\neq(\ell,\ell)$.

\bibliography{ref}

\end{document}